\documentclass[lettersize,journal]{IEEEtran}
\usepackage{amsmath,amsfonts}
\usepackage{algorithmic}
\usepackage{algorithm}
\usepackage{array}
\usepackage[caption=false,font=normalsize,labelfont=sf,textfont=sf]{subfig}
\usepackage{textcomp}
\usepackage{stfloats}
\usepackage{url}
\usepackage{verbatim}
\usepackage{multirow}
\usepackage{graphicx}

\begin{document}

\title{DS-TTS: Zero-Shot Speaker Style Adaptation from Voice Clips via Dynamic Dual-Style Feature Modulation}

\author{
    Ming Meng, Ziyi Yang, Jian Yang, Zhenjie Su, Yonggui Zhu, Zhaoxin Fan,
\thanks{Corresponding author: Zhaoxin Fan, Email: zhaoxinf@buaa.edu.cn}
\thanks{Ming Meng and Ziyi Yang contributed equally to this work.}
\thanks{Ming Meng , Ziyi Yang and Yonggui Zhu are with the School of Data Science and Media Intelligence, Communication University of China, Beijing 100049, China.}
\thanks{Jian Yang is with the Psyche AI Inc. Beijing, China.}
\thanks{Zhenjie Su is with the Hainan International College, Communication University of China,Hainan, 572400, China.}
\thanks{Zhaoxin Fan is with the Beijing Advanced Innovation Center for Future Blockchain and Privacy Computing, Institute of Artificial Intelligence, Bei hang University, Beijing, China Hangzhou International Innovation Institute, Beihang University, China.}}



\maketitle

\begin{abstract}
Recent advancements in text-to-speech (TTS) technology have increased demand for personalized audio synthesis. Zero-shot voice cloning, a specialized TTS task, aims to synthesize a target speaker's voice using only a single audio sample and arbitrary text, without prior exposure to the speaker during training. This process employs pattern recognition techniques to analyze and replicate the speaker's unique vocal features. Despite progress, challenges remain in adapting to the vocal style of unseen speakers, highlighting difficulties in generalizing TTS systems to handle diverse voices while maintaining naturalness, expressiveness, and speaker fidelity. To address the challenges of unseen speaker style adaptation, we propose DS-TTS, a novel approach aimed at enhancing the synthesis of diverse, previously unheard voices. Central to our method is a Dual-Style Encoding Network (DuSEN), where two distinct style encoders capture complementary aspects of a speaker's vocal identity. These speaker-specific style vectors are seamlessly integrated into the Dynamic Generator Network (DyGN) via a Style Gating-Film (SGF) mechanism, enabling more accurate and expressive reproduction of unseen speakers' unique vocal characteristics. In addition, we introduce a Dynamic Generator Network to tackle synthesis issues that arise with varying sentence lengths. By dynamically adapting to the length of the input, this component ensures robust performance across diverse text inputs and speaker styles, significantly improving the model’s ability to generalize to unseen speakers in a more natural and expressive manner.  Experimental evaluations on the VCTK dataset suggest that DS-TTS demonstrates superior overall performance in voice cloning tasks compared to existing state-of-the-art models, showing notable improvements in both word error rate and speaker similarity. Audio samples can be found at https://ds-tts.github.io/.
\end{abstract}


\section{Introduction}
\IEEEPARstart{I}{n} recent years, advancements in Text-to-Speech (TTS) technology \cite{4} have significantly improved the ability to convert written text into natural and fluent speech, establishing a robust foundation for diverse applications across numerous domains. In education, TTS has been employed to generate high-quality personalized audio \cite{25}, fostering tailored and adaptive learning environments. In the field of assistive technologies, progress in multilingual synthesis \cite{29} has enhanced cross-linguistic communication and optimized information accessibility for multilingual users. Similarly, in the entertainment and media industries, multi-speaker voice synthesis \cite{28} has been extensively utilized for podcast production and dynamic character voice generation, enriching user experiences.

However, alongside these advancements, the demand for personalized speech generation has escalated, particularly in the context of voice cloning. A major challenge in this area is zero-shot voice cloning \cite{32}, which requires the generation of speech in the style of an unseen speaker using only a single audio sample and arbitrary text. This task introduces significant complexity, particularly under limited data conditions, as current systems struggle to accurately replicate the unique vocal characteristics of speakers. Despite the progress in TTS, achieving high-fidelity voice cloning in zero-shot scenarios remains an open and unresolved issue in the field.

Speaker style adaptation methods fall into two main categories: speaker adaptation and speaker encoding. Speaker adaptation fine-tunes pre-trained TTS models with target speaker audio to better capture voice characteristics. For example, Adaspeech \cite{6} employs a dual encoder and conditional layer normalization, while Raviraj et al. \cite{7} optimize Tacotron2 for rapid speaker adaptation. However, these methods often suffer from overfitting when limited audio samples are available, reducing their practical effectiveness.On the other hand, speaker encoding extracts speaker-specific features from audio samples and uses them as conditional vectors for synthesis. The success of these models relies on advanced pattern recognition techniques \cite{56,57}, especially in feature representation and similarity measurement. SV2TTS \cite{8} combines a speaker verification model trained with GE2E loss with Tacotron2, and Chien et al. \cite{9} show that embeddings from speech conversion tasks yield superior results. MRMI-TTS \cite{10} incorporates a speaker-content separation module, while StyleSpeech \cite{11} uses a Mel-style encoder to effectively capture speaker style, achieving strong performance even with scarce target speaker data. Despite these advancements, the majority of models are primarily optimized for long-form sentence synthesis, where precise control over prosody and intonation is essential for achieving naturalness. Given these challenges, it is particularly necessary to explore speaker style adaptation under a zero-sample framework, which can utilize feature fusion techniques \cite{58,59} in pattern recognition to achieve fast and efficient adaptation with minimal speech data.

\begin{figure}[htbp]
    \centering
    \includegraphics[width=1\linewidth]{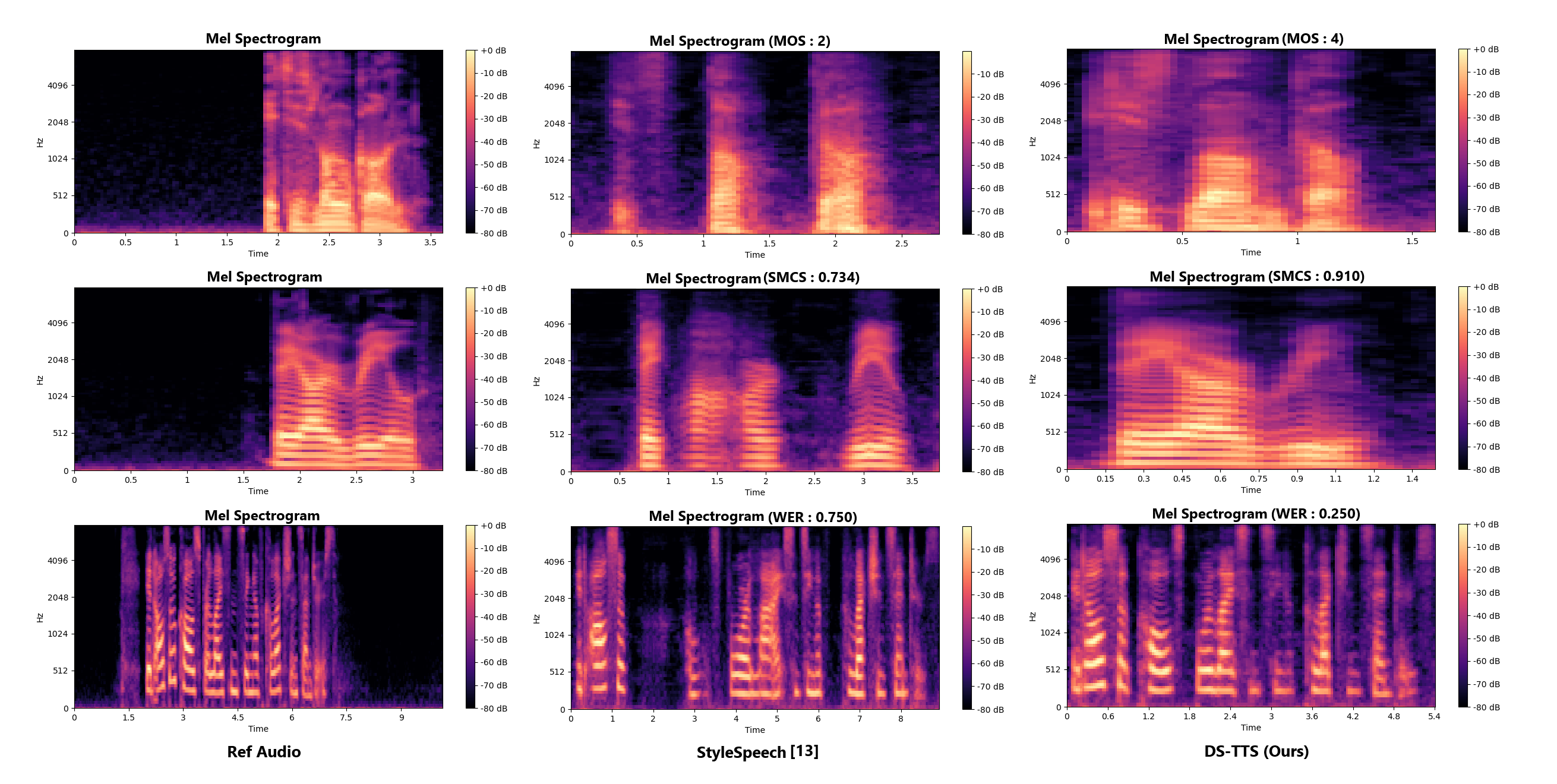}
    \caption{Comparison of mel-spectrograms of different models in voice cloning tasks. \textbf{Left:} the spectrogram of real audio. \textbf{Middle:} the synthesis results of StyleSpeech. \textbf{Right:} the synthesis results of ours model-DS-TTS.}
    \label{fig:mesh1}
\end{figure}

To this end, we propose Dynamical Style Text-to-Speech (DS-TTS), a novel approach designed to improve speaker similarity and model generalization. DS-TTS consists of two key components: the Dual-Style Encoding Network (DuSEN) and the Dynamic Generator Network (DyGN). the Dual-Style Encoding Network extracts speaker-specific style vectors from both mel-spectrograms and Mel-Frequency Cepstral Coefficients (MFCCs). These two representations capture complementary aspects of speaker identity, providing a more comprehensive style representation. The extracted vectors are then modulated using a Style Gating-Film (SGF) mechanism, which integrates these style features into the Dynamic Generator Network, enhancing speaker similarity and improving generalization to unseen speakers. To further address the challenges of synthesizing short utterances, we also introduce a dynamic variance adapter in Dynamic Generator Network. Empirical analysis shows that Convolutional Layers outperform Linear Layers for short phoneme sequences due to their ability to capture local temporal patterns. Consequently, we design two specialized predictors: one using Linear Layers for longer sequences and another using Convolutional Layers for shorter sequences. This dual approach improves the accuracy of pitch, energy, and duration predictions, ensuring robust synthesis across varying input lengths. By combining the Dual-Style Encoding Network with the dynamic variance adapter, DS-TTS effectively generalizes to unseen speakers, delivering natural and speaker-consistent audio even for challenging short utterances. As shown in Figure \ref{fig:mesh1}, DS-TTS shows significant advantages in the generated mel-spectrogram compared to existing methods (e.g., StyleSpeech \cite{11}). The spectrogram produced by DS-TTS is clearer, and the speech features are highly consistent with the reference audio. Notably, there are marked improvements in key metrics such as speech quality, speaker consistency, and audio fidelity. Moreover, extensive evaluations on the LibriTTS and VCTK datasets further validate the overall superior performance of DS-TTS in voice cloning tasks, highlighting its distinct advantages over the state-of-the-art models.

In summary, our primary contributions are as follows:
\begin{itemize}
  \item We propose DS-TTS, a multi-speaker adaptive text-to-speech (TTS) model designed to achieve high levels of speaker similarity and broad applicability. Notably, DS-TTS excels in zero-shot voice cloning tasks, enabling effective speech synthesis from previously unseen speakers without reliance on speaker-specific training data.
  \item We introduce a Dual-Style Encoder Architecture coupled with the Style Gating-Film (SGF) mechanism, which enhances the model’s ability to filter out irrelevant information and adaptively refine speaker style representations. This architecture significantly improves the accuracy of speaker style capture and the overall quality of generated speech.
  \item We develop a Dynamic Generator Network that flexibly adjusts computational resources based on input sequence length. By optimizing network complexity according to the specific requirements of short and long text sequences, this method enhances processing efficiency and improves synthesis quality across varying sentence lengths.
  \item We conduct extensive experiments on the LibriTTS and VCTK datasets to validate the effectiveness of DS-TTS. Experimental results, based on both subjective and objective metrics, demonstrate that DS-TTS showcasing superior voice cloning capabilities and overall performance.
\end{itemize}

\section{Related Work}
\subsection{Voice Cloning}
Voice cloning has witnessed significant advancements in recent years, particularly with models that leverage speaker encoding techniques \cite{42,43}. These models typically extract a style vector from reference audio, which encapsulates the speaker's unique vocal characteristics, and then incorporate this vector into the training of the text-to-speech (TTS) model. Notable approaches include Meta-StyleSpeech \cite{11}, which employs meta-learning to enhance the model's ability to produce audio with high fidelity and naturalness. Generspeech \cite{13} introduces a multi-level style adapter and a generalizable content adaptor featuring Mix-Style Layer Normalization, allowing for effective simulation of speaker styles. MRMI-TTS \cite{10} utilizes both a speaker encoder and a content encoder to derive speaker and content embeddings from multiple reference audio samples, applying mutual information minimization to achieve disentanglement. Openvoice \cite{14} enables style control while preserving the speaker's timbre. However, many existing models primarily extract features from mel-spectrograms and overlook the speaker information encoded in mel-frequency cepstral coefficients (MFCC). To address this gap, we propose a dual-style encoder framework that simultaneously leverages both mel-spectrogram and MFCC features to enhance speaker style extraction and representation.

\subsection{Conditional Sequence Generation Model}
Conditional sequence generation is a pivotal area in deep learning that focuses on leveraging conditional information to produce specific types of data. Skerry-Ryan et al. \cite{15} integrated speaker embeddings with text features and fed them into the Tacotron model, facilitating speaker style transfer. Huang et al. \cite{16} introduced adaptive instance normalization (AdaIN), which provides a straightforward and effective method for real-time style transformation. Chou et al. \cite{17} utilized AdaIN to convey speaker information to the decoder, achieving voice conversion. YourTTS \cite{18} employed a style encoder to extract style features, which were then integrated into text tensors via cascade and attention mechanisms, enabling the generation of features with distinct speaker styles. FiLM \cite{19} applies two generated parameters to scale and decode features at each layer, thereby adaptively influencing the neural network's output. Building on these advancements, our approach introduces Style Gating-Film, a novel mechanism designed to selectively filter out information that adversely affects learning. This method enhances the model's ability to generalize across diverse zero-shot data, improving the effectiveness of transferring global knowledge to different audio contexts.

\subsection{Intelligent Dynamic Neural Network}
Dynamic neural networks \cite{12} have become increasingly influential across various domains, including computer vision \cite{60,61,62} and natural language processing \cite{63,64}. For example, SkipNet \cite{21} introduces a dynamic skipping mechanism where gated networks selectively bypass certain layers. Similarly, dynamic early exit networks \cite{20,53} make routing decisions based on intermediate classifiers. Yu et al. \cite{22} propose a depth-adaptive neural network that truncates the least significant bits and adjusts the bit width during runtime to balance speed and accuracy. Park et al. \cite{23} demonstrate adaptive reasoning by cascading two deep neural networks (DNNs) with varying depths, while Li et al. \cite{24} employ cascaded convolutional neural networks (CNNs) to handle different image resolutions, improving face detection across scales. Dynamic routing methods, such as selecting and executing one of several candidate modules at each layer \cite{54,55}, have also been explored to enhance performance. The advent of dynamic neural networks provides new insights into model generalization, allowing tailored computational effort for each sample and improving model expressiveness. In the context of voice cloning, complex networks can enhance generalizability; however, simpler inputs may require only shallow networks. Consequently, this paper proposes a Dynamic Generator Network that adapts to varying phoneme sequence lengths, thereby improving the model's expressive capability.

\section{DS-TTS}

\begin{figure*}
    \centering
    \includegraphics[width=1\linewidth]{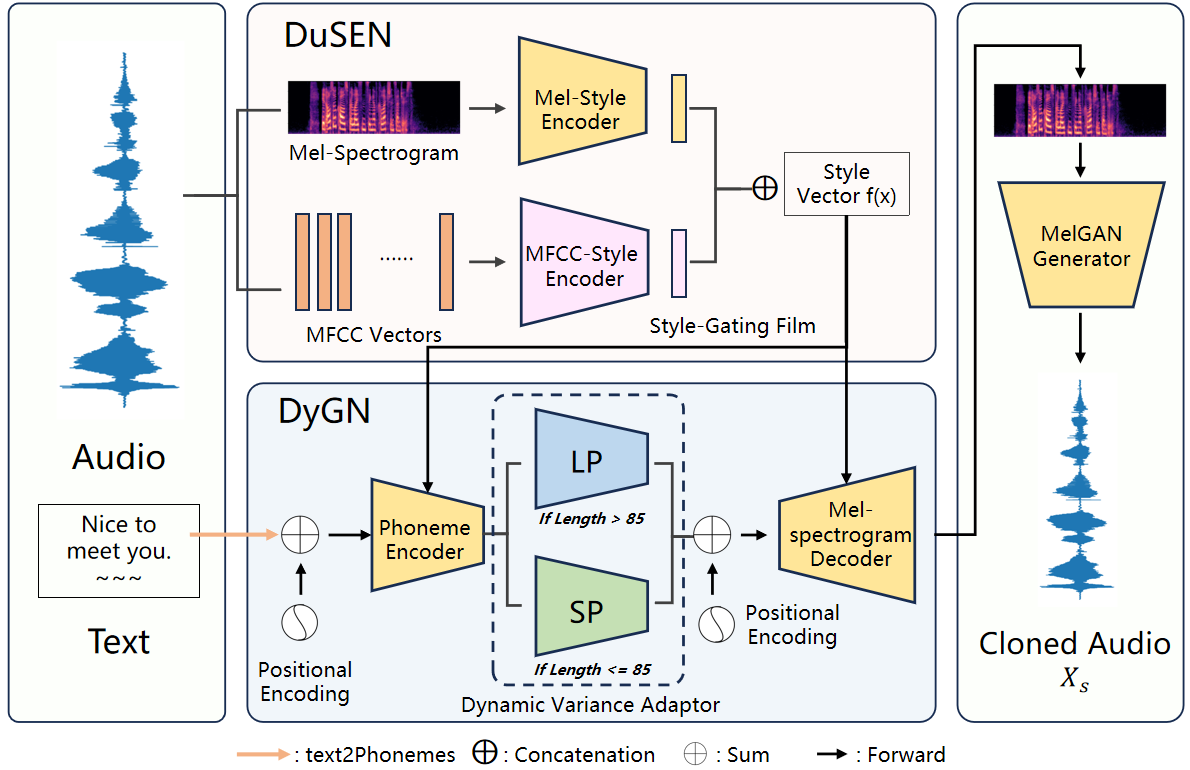}
    \caption{The overall architecture of DS-TTS, which consists of the Dual-Style Encoding Network (DuSEN) and the Dynamic Generator Network (DyGN). The Dual-Style Encoding Network extracts speaker-specific style features, while the Dynamic Generator Network synthesizes speech based on the input phonemes and extracted style vector.}
    \label{fig:mesh2}
\end{figure*}

\subsection{Problem Formulation}
In the task of zero-shot voice cloning, our objective is to construct a model \( f \) that can generate fluent and highly similar speech \( x_s \) from an unseen speaker audio sample \( x \) and a text \( t \). Specifically, given the target speaker's audio sample \( x \) and the input text \( t \), the model must ensure that the generated speech \( x_s \) is the same as the reference audio \( x \), which can be quantified using the loss function \(\mathcal{L} =\|x - x_s||\). Additionally, the generated speech \( x_s \) must maintain consistency with the target speaker's audio sample \( x \), evaluated through the similarity measure \(S(x, x_s) = \frac{\langle x, x_s \rangle}{\| x \| \| x_s \|}.\) We observe that the model performed poorly in synthesizing short sentences, with the generated speech \( x_s \) often appearing unnatural and discontinuous. This indicates that the current model architecture lacks adequate adaptability for short sentence synthesis. To address this issue, we propose a novel design that combines Dual-Style Encoding Network (DuSEN) and Dynamic Generator Network (DyGN) to optimize the model's short sentence synthesis capability. Specifically, by adjusting the input features \( \mathbf{f}(x) \) through DuSEN and introducing the dynamic variance adapter \( D(l) \) in the Dynamic generator network, the model can adjust its computational complexity based on the input sequence length \( l \). This approach aims to more accurately capture the target speaker's unique vocal characteristics, thereby enhancing the naturalness and similarity of short sentence synthesis, and improving the practical application of zero-shot voice cloning technology.

\subsection{Model Architecture Overview}
The proposed DS-TTS (Dynamical Style Text-to-Speech) model aims to address the challenge of zero-shot voice cloning, i.e. synthesizing high-quality, speaker-specific speech without ever seeing the target speaker. The model architecture consists of two main sub-networks: Dual-Style Encoding Network (DuSEN) and Dynamic Generator Network (DyGN). These two sub-networks work together to achieve robust zero-shot voice cloning by extracting speaker features from reference audio and combining them with input text to generate fluent and natural speech. As illustrated in Figure \ref{fig:mesh2}, the DS-TTS operates in two stages. In the first stage, the DuSEN leverages two distinct style encoders to extract comprehensive style vectors from reference audio. These vectors capture complementary aspects of the speaker’s vocal characteristics, ensuring that the synthesized speech closely mirrors the target speaker’s unique features. In the second stage, the DyGN synthesizes speech by dynamically adapting to varying phoneme sequence lengths and speaker styles, using both phoneme inputs and the extracted style vectors. To ensure seamless integration between the two sub-networks, the model employs a Style Gating-Film (SGF) mechanism, which injects the speaker’s style information into both the phoneme encoder and the mel-spectrogram decoder. This mechanism ensures that the speaker’s style influences the entire synthesis process, resulting in accurate and expressive speech generation. The final mel-spectrogram is converted into a waveform using a pre-trained vocoder MelGAN, producing high-quality audio output.
\subsection{Dual-Style Encoding Network}
In the proposed DS-TTS model, the Dual-Style Encoding Network (DuSEN) plays a pivotal role in capturing speaker-specific characteristics, enabling high-quality speech synthesis. The DuSEN leverages two distinct encoders: the Mel-Style encoder and the MFCC-Style encoder, which process complementary acoustic features of the input audio to generate a comprehensive speaker style vector. This vector effectively encapsulates multiple attributes of the speaker, including tone, pace, intonation, and pronunciation habits, enabling robust zero-shot voice cloning of unseen speakers.

\textbf{The input audio} is first transformed into two complementary acoustic representations: mel-spectrograms and Mel-Frequency Cepstral Coefficients (MFCCs). mel-spectrograms capture global acoustic features, such as prosody and fundamental frequency contours, representing the overall sound structure. In contrast, MFCCs focus on spectral details, aiding the model in understanding subtle nuances in vocalization characteristics and timbre. This combination provides a comprehensive description of the speaker's voice identity, ensuring that the generated speech closely matches the target speaker's natural vocalization in terms of naturalness and consistency.

\textbf{The Mel-Style encoder} extracts global acoustic features from the mel-spectrogram. The encoder produces a 128-dimensional style vector that captures essential characteristics such as pitch, rhythm, and energy, which are crucial for maintaining the naturalness and fluency of the synthesized speech. Through the Mel-Style encoder, the model can accurately reproduce the target speaker’s fundamental vocal properties.

\textbf{The MFCC-Style encoder} inspired by the design of Zou et al. \cite{33}, focuses on extracting finer speaker characteristics from the MFCC representation. This encoder consists of three main components: a Bidirectional LSTM that models the temporal dependencies and dynamics in the MFCC data, allowing the model to capture time-dependent variations in the speaker’s voice; a Multi-head Self-Attention Mechanism that calculates global contextual information from the MFCC features, enhancing the model’s understanding of long-range dependencies and complex speech nuances; and Average Pooling, which reduces the time dimension by averaging the output of the self-attention mechanism, generating a fixed-size style vector. This vector has the same dimensionality as the Mel-Style encoder’s output, facilitating the subsequent concatenation operation.

Finally, the outputs of the Mel-Style encoder and the MFCC-Style encoder are concatenated to form a 256-dimensional style vector, capturing both global and local characteristics of the speaker's voice. Even for unseen speakers during training, this comprehensive style vector enables the model to accurately reproduce the target speaker's unique voice characteristics, achieving robust zero-shot voice cloning.

\subsection{Dynamic Generator Network}
In the proposed DS-TTS framework, the Dynamic Generator Network (DyGN) emerges as a pivotal component, tasked with predicting crucial acoustic features, including pitch, energy, and duration. Crucially, DyGN dynamically adapts to varying phoneme sequence lengths, ensuring that the synthesized speech retains its natural expressiveness, a paramount consideration in zero-shot voice cloning tasks where input sequences can exhibit significant variability. At the core of DyGN lies a tripartite architecture comprising a phoneme encoder, a variance adaptor, and a mel-spectrogram decoder. These modules synergistically integrate the speaker's style vector into the synthesis process, yielding highly naturalistic and personalized speech outputs. 
\begin{figure}
    \centering
    \includegraphics[width=1\linewidth]{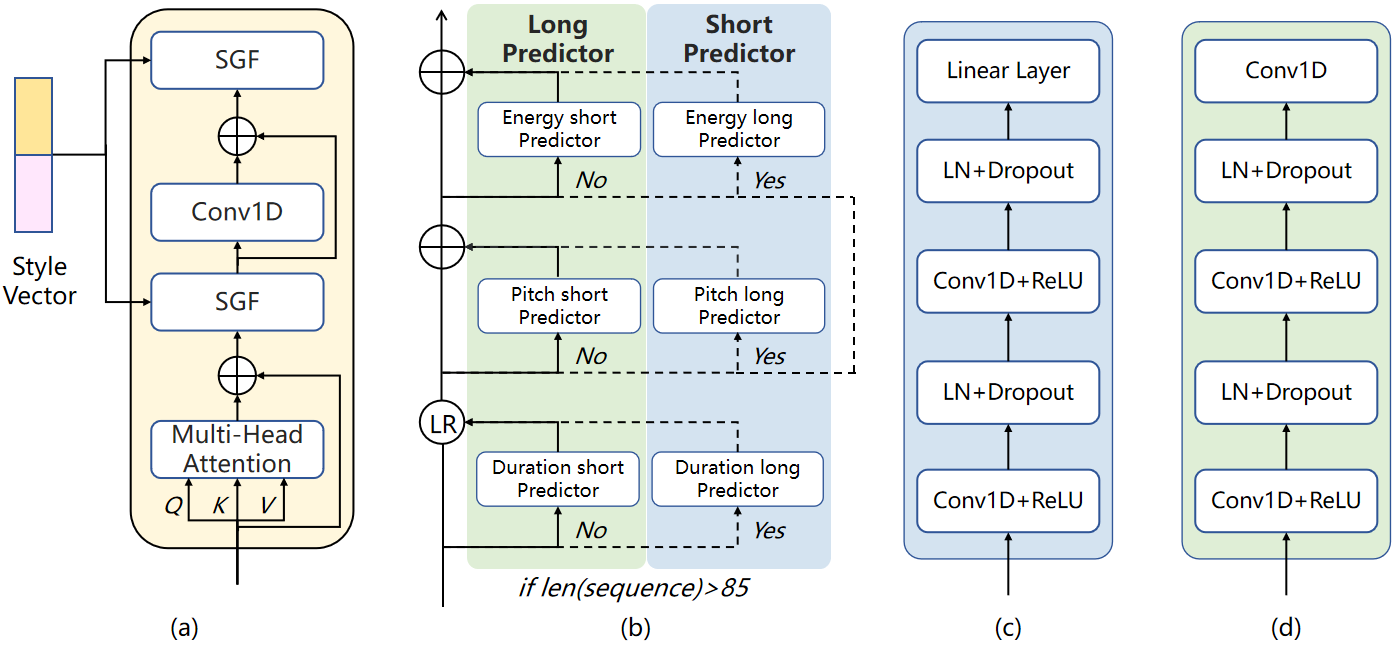}
    \caption{Details of some components. (a)FFT Block. (b) Variance Adaptor. (c)Duration/Pitch/Energy long Predictor. (d)Duration/Pitch/Energy short Predictor.}
    \label{fig:mesh3}
\end{figure}

\textbf{Phoneme Encoder} is capable of converting phoneme embedding sequences into hidden phoneme sequences. Using the Feed-Forward Transformer blocks (FFT), as shown in Figure \ref{fig:mesh3}(a), it maps the raw phoneme embedding sequence into hidden representations that effectively capture the distinctive speech style characteristics of the target speaker. Simultaneously, the speaker style vector generated by the DuSEN is injected into the FFT to modulate these hidden representations.

Traditionally, speaker style vectors can be fused with the phoneme sequence's feature information through methods such as concatenation, direct addition, or Style-Adaptive Layer Norm (SALN) \cite{11}. However, concatenation and direct addition can lead to confusion in semantic information and fail to accurately convey the speaker’s fine-grained style variations. Although SALN offers an adaptive way to scale and shift feature vectors based on the style vector, it lacks the ability to filter out detrimental information that may negatively impact learning. To address these limitations, inspired by the TaNP model \cite{34}, we introduce the Style Gating-Film (SGF) mechanism to refine the modulation of hidden features. SGF accepts the speaker style vector \( \mathbf{f}(x) \) and modulates the statistical distribution of hidden feature representations. It introduces two additional modulation parameters, \(\eta\) and \(\delta\), to provide finer control over the mean and variance of the feature vectors. Specifically, the feature vector \( h = (h_1, h_2, \dots, h_N) \), where \(N\) is the dimension of the vector, is first normalized:
\[
y = \frac{h - \mu}{\sigma}
\]
\[
\mu = \frac{1}{N} \sum_{i=1}^{N} h_i \quad , \quad \sigma = \sqrt{\frac{1}{N} \sum_{i=1}^{N} (h_i - \mu)^2}
\]

Subsequently, the style vector modulates the normalized feature information using the following parameters:
\[
\gamma = \tanh(\mathbf{f}(x)), \quad \beta = \tanh(\mathbf{f}(x)), \quad \eta = \tanh(\mathbf{f}(x)), \quad \delta = \sigma(\mathbf{f}(x))
\]

The modulated parameters are then applied via a weighted operation to control the fine-grained stylization of the input features:
\[
\gamma = \gamma \cdot \delta + \eta \cdot (1 - \delta)
\]
\[
\beta = \beta \cdot \delta + \eta \cdot (1 - \delta)
\]

Finally, the modulated output is obtained through:
\[
SGF(y) = \gamma \cdot y + \beta
\]

This mechanism allows for precise scaling and shifting of the normalized features through the parameters \(\beta\) and \(\gamma\), thereby enabling adaptive adjustment of the features. Additionally, \(\eta\) and \(\delta\) introduce further flexibility, balancing the modulation process with two distinct parameters to support more granular semantic representation and style conversion. This ensures that the personalized style of the speaker is faithfully preserved and reproduced in the subsequent speech synthesis stages, maintaining high fidelity in zero-shot voice cloning tasks.

\textbf{Dynamic Variance Adaptor}, as a core component of the DyGN framework, plays a crucial role in dynamically predicting and adjusting key acoustic features, such as pitch, energy, and duration. By flexibly handling phoneme sequences of varying lengths, it generates speech that is both personalized and naturally fluent. This design is particularly well-suited for zero-shot voice cloning tasks, ensuring that the generated speech not only retains the unique vocal characteristics of the target speaker but also maintains high quality and naturalness throughout the synthesis process.

To handle phoneme sequences of different lengths, the DVA module incorporates a dynamic scheduling mechanism, as shown in Figure \ref{fig:mesh3}(b), which divides the phoneme sequences into short sequences ($\leq 85$ phonemes) and long sequences ($>85$ phonemes), and designs different predictor architectures for them. Each predictor is responsible for the modeling tasks of pitch, energy, and duration. The pitch predictor flexibly selects short or long sequence predictors according to the length of the phoneme sequence, and uses a convolutional layer (Conv1D) or a linear layer to model the pitch respectively; the energy predictor is responsible for predicting the energy value of each phoneme to ensure the naturalness and expressiveness of the generated speech; the duration predictor estimates the duration of the phoneme to ensure that the synthesized speech is highly consistent with the target speaker’s natural speech rate. Specifically, short phoneme sequences are processed by Conv1D, which can effectively capture local fine-grained features and avoid model overfitting due to its local connection and weight sharing characteristics, as shown in Figure \ref{fig:mesh3}(c). Long phoneme sequences, on the other hand, are handled by linear layers, which have greater expressive power and can handle more complex phoneme sequence mappings, as illustrated in Figure \ref{fig:mesh3}(d).

During training, the DVA module employs the Mean Squared Error (MSE) loss function to calculate the difference between predicted\(\ z' \) and true values\(\ z \), as shown in the following formula:
\[
MSE(z,z') = \frac{1}{n} \sum_{i=1}^{n} (z_i - z'_i)^2
\]
By optimizing this loss function, the DVA module can significantly enhance the overall performance of the model, especially when processing short phoneme sequences, where convolutional layers are used to prevent overfitting. This design not only improves the overall quality of speech synthesis but also enhances the model’s robustness across various input scenarios, ensuring the synthesized speech achieves high levels of naturalness and fidelity.

\textbf{Mel-Spectrogram Decoder} gradually generates high-resolution mel-spectrograms by receiving hidden feature sequences from the variance adapter, which contain key acoustic features such as pitch, energy, and duration. As an intermediate representation of speech signals, mel-spectrograms can effectively capture the changes in speech in the time and frequency domains, ensuring that the generated speech output is highly natural and high-quality. This decoder accurately preserves the unique characteristics of the speaker, so that the final speech synthesis not only performs well, but also accurately restores the personalized style of the target speaker. It utilizes a combination of feed-forward networks, self-attention mechanisms, and convolutional layers. The self-attention mechanism excels at modeling long-range dependencies in speech signals, ensuring that the generated speech is coherent. Meanwhile, the convolutional layers focus on capturing local features, ensuring that details are accurately rendered. By combining global and local features, the decoder is able to ensure that acoustic properties such as pitch, energy, and duration are accurately represented during synthesis.
\section{Experiment Setup}
\subsection{Dataset}
In the training phase, we select clean and noise-free subsets of the LibriTTS dataset \cite{65}, specifically the train-clean-100 and train-clean-360 subsets, which collectively encompass speech samples from 1,148 speakers. Additionally, to assess the model's generalization capability with previously unseen speakers, we incorporated the VCTK dataset, which contains speech data from 108 speakers. Prior to model training, a series of preprocessing steps were implemented on both the audio and text data to ensure data quality and consistency. First, the sampling rate for all training and evaluation data was uniformly converted to 16 kHz. Subsequently, we employed Fast Fourier Transform (FFT) techniques to extract spectrograms, setting the window size to 1024 samples and the hop size to 256 samples, ultimately converting the data into mel-spectrograms with 80 frequency bins. Next, we used G2P to convert text into phoneme sequence, and sutilized the librosa library to extract 20 Mel-frequency cepstral coefficients (MFCCs) from the mel-spectrograms, applying L2 norm normalization to enhance feature comparability. Finally, the alignment of audio data with text data was achieved using the Montreal Forced Aligner, enabling the accurate duration extraction for each phoneme. This systematic preprocessing workflow provides high-quality training and evaluation data for the subsequent zero-shot voice cloning task, ensuring the model's effectiveness and robustness across diverse speaker samples.

\subsection{Implementation Details}
The proposed model comprises five key modules: the Mel Style Encoder, the MFCC Style Encoder, the Phoneme Encoder, the Dynamic Variance Adapter, and the Mel-Spectrogram Decoder. For audio synthesis, we employ the \textit{MelGAN\footnote{\url{https://huggingface.co/Guan-Ting/StyleSpeech-MelGAN-vocoder-16kHz}}} vocoder \cite{38}, which converts mel-spectrograms into audible audio waveforms. Both the Mel Style Encoder and the MFCC Style Encoder produce 128-dimensional output vectors, which are concatenated to form a 256-dimensional style vector. To facilitate the integration of style information, a Style Gating-FiLM layer is implemented within the FFT of both the Phoneme Encoder and the Mel-Spectrogram Decoder. During the initialization phase, we set the bias for the parameter $\delta$ to 1.0, while the biases for the other parameters $\gamma$, $\beta$ and $\eta$ are initialized to 0. This design choice aims to ensure a balanced model state at the onset, thereby optimizing its adaptability to subsequent style modulation. In the Dynamic Variance Adapter, the architecture for predicting pitch, energy, and duration for longer phoneme sequences consists of two layers of 1D convolutional networks with ReLU activations, each followed by normalization and dropout layers, culminating in a linear layer as the output. For shorter phoneme sequences, the predictor is similarly structured but utilizes a convolutional layer as the output instead of a linear layer. The training batch size for the model is set to 24, utilizing the Adam optimizer with parameters $\beta_1=0.9$, $\beta_2=0.98$, $\epsilon=10^-9$.

In the training process, the model is optimized by minimizing the mel-spectrogram reconstruction loss, duration prediction loss, energy prediction loss, and pitch prediction loss. Specifically, the mean absolute error (MAE) loss function is used to calculate the reconstruction loss of the mel-spectrogram, while MSE loss function is used to calculate the losses of other features. The total loss function is calculated as follows:
\[
Loss = L_{rec} + L_d + L_e + L_p
\]
where $L_{rec}$ represents the mel-spectrogram reconstruction loss, $L_d$ represents the duration prediction loss, $L_e$ represents the energy prediction loss, and $L_p$ represents the pitch prediction loss. The model undergoes training for 200,000 steps, ensuring effective performance for zero-shot voice cloning tasks.

\subsection{Baselines}
In this study, we meticulously design the implementation details of the model to ensure the reliability and validity of the experiments. To comprehensively evaluate the proposed DS-TTS model, we employ multiple baseline methods for comparison, ensuring the fairness and scientific rigor of the evaluation process. Additionally, we conduct ablation experiments to gain deeper insights into the impact of each component on model performance.
\newline
\textbf{Model Comparison and Evaluation.} All comparative methods were assessed using their official code and pre-trained checkpoints to ensure a systematic evaluation of performance. Given that the VALL-E-X model synthesizes speech at a sampling rate of 24,000 Hz, we downsampled all generated audio to this rate during the evaluation process, while maintaining a sampling rate of 16,000 Hz for other baseline models. This strategy ensures consistency and comparability of the evaluation results. The baseline models for comparison include: (1) \emph{Ground Truth (GT)} represents real speech samples and serves as the benchmark for performance evaluation in this study, providing an ideal reference. (2) \emph{StyleSpeech\footnote{\url{https://github.com/KevinMIN95/StyleSpeech}}} utilizes a Mel-Style encoder to extract speaker style features, integrating these features into a FastSpeech2-based speech synthesis framework through a style-adaptive layer normalization, thereby generating audio for the target speaker. (3) \emph{Retrained StyleSpeech} involves adjusting the batch size to 24 and retraining the StyleSpeech model on the same dataset partition for 200,000 steps. This approach not only ensured the model was properly trained but also enhanced the accuracy of the performance comparisons, facilitating a fair assessment of its effectiveness. (4) \emph{YourTTS \cite{43}\footnote{\url{https://github.com/coqui-ai/TTS}}} is based on VITS and employs speaker consistency loss as its objective function. We evaluated its performance using checkpoints from Coqui TTS to ensure the validity of the method. (5) \emph{VALL-E-X \cite{40}\footnote{\url{https://github.com/Plachtaa/VALL-E-X?tab=readme-ov-file}}} is based on VALL-E \cite{39} and is a cross-language neural encoder-decoder that uses reference audio and text as cues to predict the acoustic token sequences of the target audio, demonstrating its potential for cross-language applications. (6) \emph{Coqui AI XTTS v2\textsuperscript{3}}\cite{41}uses a Perceiver model, which serves as a prefix for the GPT decoder to achieve high-quality TTS synthesis. (7) \emph{StyleTTS2 \cite{42}\footnote{\url{https://github.com/sidharthrajaram/StyleTTS2}}} uses style diffusion and adversarial training with large speech language models to achieve high-quality TTS synthesis.
\newline
\textbf{Ablation Study.} To gain a deeper understanding of the impact of various model components on performance, we conducted several ablation experiments, which include the following: (1) \emph{DS-TTS w/o MFCC}: This model uses only the Mel-Style encoder to extract speaker embedding features, removing the MFCC features to evaluate the contribution of MFCC to the overall model performance. (2) \emph{DS-TTS w/o DVA-SP:} In this variant, the short predictors (DVA-SP) in the Dynamic Variance Adapter is removed to investigate the role of the short predictors in the quality of the generated audio. (3) \emph{DS-TTS w/o DVA-LP:} In this variant, the long predictors (DVA-LP) in the Dynamic Variance Adapter is removed to investigate the role of the long predictors in the quality of the generated audio. Through these ablation experiments, we systematically analyze the contribution of each component to the model's performance, providing valuable insights for future model optimization.
\subsection{Evaluation Metrics}
To comprehensively assess the naturalness and similarity of synthesized utterances from target speakers, we employed both subjective and objective evaluation metrics.
\newline
\textbf{Subjective metrics} consist of the Mean Opinion Score (MOS) and the Similarity Mean Opinion Score (SMOS) to evaluate the quality of audio synthesis. These metrics are rated on a scale from 1 to 5, with 5 representing the best performance. MOS is used to assess listeners' evaluations of audio clarity and sentence fluency in subjective listening tests, focusing on the presence of background noise, as well as the naturalness and fluency of prosody, including variations in intonation, pitch, and rhythm. SMOS evaluates listeners' subjective perceptions of the similarity between synthesized audio and reference audio.
\newline
\textbf{Objective metrics} include the Word Error Rate (WER) and the Speaker Embedding Cosine Similarity (SMCS) to quantify audio synthesis quality. WER measures the difference between the reference text and the text recognized by an Automatic Speech Recognition (ASR) model. A lower WER indicates greater accuracy in the pronunciation of synthesized audio. We selected the base model from the Whisper library as our ASR model. SMCS is used to compare the cosine similarity between synthesized audio and ground truth audio, with higher values indicating greater similarity in speaker voice characteristics within the synthesized audio.  We utilize \textit{Resemblyzer\footnote{\url{https://github.com/resemble-ai/Resemblyzer}}} to extract speaker representations.

\section{Results and Analysis}
In this section, we present the results and analysis of our proposed DS-TTS model. We begin by comparing the performance of DS-TTS against various baseline methods, providing a comprehensive evaluation of its effectiveness in zero-shot voice cloning. Following this, we delve into ablation studies that investigate the individual contributions of different components within the DS-TTS architecture.
\subsection{Comparison Results}
In the comprehensive evaluation of the DS-TTS model, we focus on its comparative performance across three key dimensions. First, we assess the model's effectiveness in synthesizing audio for unseen speakers to verify its generalization capability. Second, we analyze the synthesis quality of short and long phoneme sequences, exploring the model's adaptability to variations in input length. Finally, we investigate the influence of gender and accent on the synthesis results to gain a deeper understanding of the model's performance across different vocal characteristics. 

\begin{table}[ht]
\begin{center}
\caption{Evaluation of zero-shot voice cloning on unseen speakers using different models. Reporting the Word Error Rate (WER, with lower values indicating better performance) and Speaker Embedding Cosine Similarity (SMCS, with higher values indicating better speaker similarity) for each method.}
\label{tab:1}
 \begin{tabular}{c c c} 
 \hline
   Methods & WER↓ & SMCS↑  \\ [0.5ex] 
 \hline
 StyleSpeech \cite{11} & 0.064 & 0.797 \\
 Retrained StyleSpeech & 0.051 & 0.812  \\
 YourTTS \cite{43} & 0.093 & 0.785  \\
 VALL-E-X \cite{40}& 0.228 & \textbf{0.868} \\
 StyleTTS2 \cite{42}& 0.039 & 0.841  \\
 Coqui AI XTTS v2 \cite{41} & \textbf{0.032} & 0.803  \\
 DS-TTS (Ours)& 0.047 & \textbf{0.865}  \\
 \hline
\end{tabular}
\end{center}
\end{table}

\textbf{Evaluation on Unseen Speakers.} We randomly select 108 audio samples from different speakers from the VCTK dataset as the test set to evaluate the model's ability to generalize to unseen speakers. Table \ref{tab:1} shows the results of WER and SMCS. The results show that the DS-TTS model  performs exceptionally across various metrics. While the WER of DS-TTS is slightly higher than that of StyleTTS2 and XTTS v2, its similarity performance is markedly superior. Although the SMCS of VALL-E-X is comparable to that of DS-TTS, its WER is notably higher, at 0.228. This result is related to the limited training data utilized by the authors, who employed only 704 hours of English training data, in contrast to the approximately 60,000 hours of unannotated speech data available in the LibriLight dataset. Overall, DS-TTS demonstrates impressive speaker similarity performance, alongside a low word error rate and high-quality speech synthesis capabilities.

\textbf{Evaluation of Short and Long Phoneme Sequence.} To further assess the performance of our model, we conduct baseline tests on both short and long phoneme sequences. The results, presented in Table \ref{tab:2}, include both objective and subjective evaluation metrics. First, the comparison between long and short phoneme sequences indicates that the DS-TTS model demonstrates excellent adaptability when processing reference speech of varying lengths. Second, according to the objective metrics, DS-TTS achieves the best balance between WER and SMCS among all models, highlighting its superior performance. Finally, subjective evaluation results reveal that under the same small-batch training conditions, the audio generated by DS-TTS outperforms YourTTS, StyleSpeech, and Retrained StyleSpeech in terms of MOS scores. However, when compared to StyleTTS2 and XTTS v2, DS-TTS's performance is slightly lower. This discrepancy is understandable, given that both StyleTTS2 and XTTS v2 used exceptionally large datasets, and their training times were significantly longer than ours. Due to these substantial differences in data volume and training duration, it is expected that these models would surpass ours in terms of audio generation quality. Nevertheless, it is worth noting that DS-TTS exhibited remarkable generalization capability in the subjective similarity evaluation, producing audio highly similar to that of the target speaker, demonstrating its effectiveness and potential in zero-shot voice cloning tasks.

\begin{table*}[ht]
\centering
\caption{Performance evaluation on short and long phoneme sequences across different models. Reporting the Word Error Rate (WER, lower is better) and Speaker Embedding Cosine Similarity (SMCS, higher is better) for both short sequences (length  $\leq 85$) and long sequences (length $> 85$). Higher mean opinion score (MOS) and similarity mean opinion score (SMOS) scores indicate better performance.}
\label{tab:2}    
\resizebox{\textwidth}{!}{%
\begin{tabular}{c c c c c|c c c c}
 \hline
 \multirow{2}{*}{} & \multicolumn{4}{c|}{Short (len $ \leq 85 $)} & \multicolumn{4}{c}{Long (len $>$ 85)} \\
 \cline{2-5} \cline{6-9}
Methods & WER↓ & SMCS↑ & MOS↑ & SMOS↑ & WER↓ & SMCS↑ & MOS↑ & SMOS↑ \\
 \hline
 GT & - & - & 4.42 & - & - & - & 4.41& - \\ 
 StyleSpeech \cite{11} & 0.064 & 0.793	& 3.83 & 3.58 & 0.022 & 0.827 & 4.04 & 3.95 \\
 Retrained StyleSpeech & 0.058 & 0.808 & 3.72 & 3.44 & 0.012 & 0.854 & 4.07 & 3.86   \\
 YourTTS \cite{43}& 0.124	& 0.785	& 3.95 & 3.57 & 0.057 & 0.838 & 4.01 & 3.67   \\
 VALL-E-X \cite{40}& 0.218 & \textbf{0.863} & 4.09 & 3.48 & 0.339 & \textbf{0.913} & 3.96 & 3.57 \\
 StyleTTS2 \cite{42}& 0.045 & 0.839 & 4.21 & 3.46  & 0.026 & 0.884 & \textbf{4.32} & 3.74   \\
 Coqui AI XTTS v2 \cite{41} & \textbf{0.030} & 0.810 & \textbf{4.23} & 3.59  & 0.019 & 0.856 & 4.29 & 3.56  \\
 DS-TTS (Ours)& 0.052 & \textbf{0.863} & 4.08 & \textbf{3.72}  & \textbf{0.011} & 0.868 & 4.12 & \textbf{3.99}  \\
 \hline
\end{tabular}
}
\end{table*}

\begin{table}[ht]
\begin{center}
\caption{Evaluation of DS-TTS model performance across different genders and English accents. Reporting the Word Error Rate (WER, lower is better) and Speaker Embedding Cosine Similarity (SMCS, higher is better) for male and female speakers, as well as for various English accents.}
\label{tab:3}
\begin{tabular}{ c c c c }
\hline
Metric & & WER↓ & SMCS↑ \\
\hline
\multirow{2}{*}{Gender} & Male & 0.050 & 0.862 \\
& Female & 0.037 & 0.864 \\
\hline
\multirow{5}{*}{Accent} & American & 0.040 & 0.867 \\
& Canadian & 0.046 & 0.859 \\
& Irish & 0.046 & 0.866 \\
& NorthernIrish & 0.051 & 0.856 \\
& Scottish & 0.048 & 0.859 \\
\hline
\end{tabular}
\end{center}
\end{table}

\textbf{Evaluate Gender and Accent.} We select 100 audio samples for each category as the test set in the evaluation of different genders and English accents. The results are shown in Table \ref{tab:3}. The DS-TTS model is able to accurately generate audio that is highly similar to the original speech, and there is no obvious difference between male and female speakers. In addition, the model is able to effectively capture the characteristics of various English accents with high accuracy. This shows that the feature extraction method used in this study successfully captures the commonalities and differences between different genders and accents, thereby ensuring balanced performance in the synthesis process. These results further demonstrate the robustness and generalization ability of DS-TTS in handling various speech synthesis tasks.

\subsection{Ablation Studies for DS-TTS}
\subsubsection{Ablation Studies Were Performed on Thresholds}
The selection of an appropriate threshold for phoneme sequence length significantly impacts the overall performance of the DS-TTS model. To assess this, ablation studies are conducted by evaluating the model across a range of thresholds from 75 to 95, as shown in Table \ref{tab:4}. The performance is measured using WER and SMCS, which assesses how well the synthesized speech matches the target speaker's voice.

\begin{table}[ht]
\begin{center}
\caption{Ablation study results for different phoneme sequence length thresholds (75 to 95). Reporting the Word Error Rate (WER, with lower values indicating better performance)
and Speaker Embedding Cosine Similarity (SMCS, with higher values indicating better
speaker similarity) for each threshold.}
\label{tab:4}
 \begin{tabular}{c c c} 
 \hline
  Threshold value & WER↓ & SMCS↑ \\ [0.5ex] 
 \hline
  95 & 0.059 & 0.845 \\ 
  90 & 0.067 & 0.850 \\
  85 & \textbf{0.047} & \textbf{0.865} \\
  80 & 0.051 & 0.842 \\
  75 & 0.053 & 0.838 \\
 \hline
\end{tabular}
\end{center}
\end{table}

The results indicate that the model with a threshold of 85 outperformed other configurations, achieving the lowest WER of 0.047 and the highest SMCS score of 0.865. This suggests that a threshold of 85 provides an optimal balance between accurate pronunciation and speaker similarity. The improvement can be attributed to the model's ability to efficiently process phoneme sequences of varying lengths without either truncating important features or introducing unnecessary computational complexity. Models with lower thresholds (e.g., 75 and 80) exhibited slightly lower WER but compromised on speaker similarity, whereas models with higher thresholds (e.g., 90 and 95) had improved SMCS but performed worse in terms of WER.

\subsubsection{Verify The Utility of Model Components}

We conduct ablation studies to evaluate the significance of key components in the DS-TTS model, specifically the MFCC features and the Dynamic Variance Adaptor (DVA). The results, as shown in Table \ref{tab:5}, highlight the impact of removing these components on the overall model performance in terms of WER and SMCS.

\begin{table}[ht]
\begin{center}
\caption{Ablation studies evaluating the contribution of different model components. Reporting the Word Error Rate (WER, with lower values indicating better performance)
and Speaker Embedding Cosine Similarity (SMCS, with higher values indicating better
speaker similarity) for each threshold.}
\label{tab:5}
 \begin{tabular}{c c c} 
 \hline
  Methods  & WER↓ & SMCS↑ \\ [0.5ex] 
 \hline
 DS-TTS & \textbf{0.047} & \textbf{0.865} \\ 
 DS-TTS w/o mfcc & 0.060 & 0.832 \\
 DS-TTS w/o DVA-SP & 0.066 & 0.821 \\
 DS-TTS w/o DVA-LP & 0.067 & 0.838 \\
 \hline
\end{tabular}
\end{center}
\end{table}

First, the removal of the MFCC features from the DS-TTS model results in a notable degradation in performance, with the WER increasing from 0.047 to 0.060 and the SMCS decreasing from 0.865 to 0.832. This demonstrates that MFCC features play an essential role in capturing fine-grained speaker characteristics, which are crucial for achieving high-quality speech synthesis and maintaining speaker similarity in zero-shot voice cloning tasks. Next, we evaluate the effect of omitting the short and long phoneme sequence predictors in the Dynamic Variance Adaptor (DVA-SP and DVA-LP, respectively). Both variations show a significant drop in performance compared to the full DS-TTS model. For example, when only the short phoneme predictor (DVA-SP) is removed, the WER increases to 0.066, and the SMCS drops to 0.821. Similarly, when the long phoneme predictor (DVA-LP) is omitted, the WER rises to 0.067, and the SMCS reduces to 0.838. These results suggest that both short and long phoneme predictors contribute substantially to the model's ability to handle varying phoneme sequence lengths, and their combined use is essential for accurate acoustic feature prediction.

\section{Conclusion}
This study proposes a new zero-shot voice cloning model, DS-TTS, aimed at addressing the challenge of adapting to unseen speaker styles. By introducing the Dual-Style Encoding Network (DuSEN) and Dynamic Generator Network (DyGN), we effectively capture and reproduce the unique vocal characteristics of unseen speakers, while improving the model’s adaptability and performance stability. In the Dual-Style Encoding Network, we use a Mel-Style encoder and an MFCC-Style encoder to extract the speaker's style vectors, which are then fed into the phoneme encoder and the mel-spectrogram decoder in DyGN through the Style Gating-Film (SGF) mechanism. In DyGN, we dynamically adjust to the length of the input, eabling the model to synthesize audio from unseen speakers in a more natural and expressive manner.  Experimental results show that DS-TTS significantly outperforms existing methods across multiple evaluation metrics, demonstrating its vast potential for practical applications.
\newline
\textbf{Limitations and Future Work.} Despite the notable advancements made in zero-shot voice cloning, several limitations remain. First, the performance of the DS-TTS model when handling extreme speech characteristics, such as dialects or speech with specific emotional tones, requires further validation. Second, the synthesis quality in complex background noise environments still needs improvement, posing challenges to its robustness in real-world applications. Future research will address these limitations by incorporating more complex background noise modeling and diversifying the speech samples used for training, thereby enhancing the model's adaptability and stability. Additionally, we plan to explore the integration of other deep learning architectures, such as Generative Adversarial Networks (GANs), to improve the naturalness and authenticity of synthesized audio. Future efforts will also focus on combining pre-trained models with existing encoders for joint training, further enhancing the extraction of speaker styles to improve the similarity and practical value of DS-TTS.
\bibliographystyle{IEEEtran}       
\bibliography{ref}       

\vfill
\end{document}